\documentclass[twocolumn,groupedaddress,secnumarabic]{revtex4}

\usepackage{graphicx}

\begin{document}




\title{Electrode Polarization Effects in Broadband Dielectric Spectroscopy}

\author{S. Emmert, M. Wolf, R. Gulich, S. Krohns, S. Kastner, P.
Lunkenheimer, A. Loidl}

\affiliation{Experimental Physics V, Center for Electronic
Correlations and Magnetism, University of Augsburg, 86135 Augsburg,
Germany}

\begin{abstract}

In the present work, we provide broadband dielectric spectra showing
strong electrode polarization effects for various materials,
belonging to very different material classes. This includes both
ionic and electronic conductors as, e.g., salt solutions, ionic
liquids, human blood, and colossal-dielectric-constant materials.
These data are intended to provide a broad data base enabling a
critical test of the validity of phenomenological and microscopic
models for electrode polarization. In the present work, the results
are analyzed using a simple phenomenological equivalent-circuit
description, involving a distributed parallel RC circuit element for
the modeling of the weakly conducting regions close to the
electrodes. Excellent fits of the experimental data are achieved in
this way, demonstrating the universal applicability of this
approach. In the investigated ionically conducting materials, we
find the universal appearance of a second dispersion region due to
electrode polarization, which is only revealed if measuring down to
sufficiently low frequencies. This indicates the presence of a
second charge-transport process in ionic conductors with blocking
electrodes.

\end{abstract}

\maketitle

\section{Introduction}
\label{intro} Electrode polarization (EP) leading to blocking
electrodes (BEs) is a frequently encountered phenomenon when
investigating ionic conductors with impedance or dielectric
spectroscopy. This effect, giving rise to giant values of the
dielectric constant and a strong drop of conductivity towards low
frequencies, arises when the ions arrive at the metallic electrodes
and accumulate in thin layers immediately beneath the sample surface
forming a so-called space-charge region. It occurs in solid-state
electrolytes as well as in aqueous solutions, ionic liquids, and in
many biological systems (e.g., \cite{Schwan1968a, Bauerle1969,
SCHEIDER1975, LIU1985, Nyikos1985, Bates1988, Macdonald1987,
Pimenov1996, Pimenov1998, Bordi2001, Feldman2003, Bazant2004,
Klein2006, Sanabria2006, Serghei2009, Dyre2009, Gulich2008,
Wolf2011}). EP in ionic conductors often hampers the determination
of the intrinsic dielectric properties but it also is of high
technical relevance, e.g., in the design of fuel cells or double
layer capacitors.

EP also plays an important role in the dielectric spectroscopy of
electronic conductors (e.g., \cite{LUNKENHEIMER1992, BIDAULT1994,
Lunkenheimer1996, Seeger1999, Ritus2002, Lunkenheimer2002,
Lunkenheimer2004, Biskup2005, Wang2006, Deng2007, Krohns2008,
Ferrarelli2009, Sebald2010}). In semiconducting samples it is
caused, e.g., by the formation of Schottky diodes at the
electrode-semiconductor interface. In this case the blocking of the
electrode arises from the depletion layer at the semiconductor/metal
contact in the blocked state of the diode. Recently these effects
came into the focus of interest due to their presumed importance in
the so-called "colossal dielectric constant" (CDC) materials
\cite{Lunkenheimer2002, Lunkenheimer2004, Wang2006, Deng2007,
Ferrarelli2009, Sebald2010, Lunkenheimer2010}.

The many suggestions made to account for electrode effects in
dielectric measurements can be classified into two categories,
namely, (i.) avoiding the electrode polarization by specialized
measurement techniques or (ii.) modeling these contributions, e.g.,
by equivalent circuits. The first approach, using, for example,
four-electrode methods or varying the electrode distances (e.g.,
\cite{Schwan1968a, Springer1983, Mazzeo2007}), quite generally
implies considerable additional experimental efforts compared to
standard dielectric measurements. Thus, often the second approach is
adopted and numerous, partly quite sophisticated phenomenological
and microscopic models have been proposed to model EP effects in
dielectric spectra (e.g., \cite{SCHEIDER1975, LIU1985, Nyikos1985,
Bates1988, Macdonald1987, Bazant2004, Klein2006, Sanabria2006,
Serghei2009, Dyre2009, Lunkenheimer2002, Fricke1932, Chang1952,
Macdonald1953, Franceschetti1979, Feldman1998, Macdonald2002,
Macdonald2010, Macdonald2011, Bazant2011}). In dielectric studies,
one often deals with situations where the detected EP effects are
not in the focus of interest but still cannot be ignored since they
have to be taken into account for an unequivocal determination of
the intrinsic properties. In such cases, for a straightforward
modeling of EP effects, simple equivalent circuits are usually
employed, which are assumed to be connected in series to the
intrinsic bulk contribution of the sample material. The most common
ones are a parallel RC circuit \cite{Dyre2009, LUNKENHEIMER1992,
Seeger1999, Ritus2002, Schwan1968, Umino2002} or a so-called
constant phase element (CPE) \cite{LIU1985, Bates1988,
Macdonald1987, Bordi2001, Fricke1932, Feldman1998,
Bottelberghs1976}, the latter being mainly applied for modeling
ionic conductors. Both models work relatively well if only the onset
of EP effects is seen, arising close to the low-frequency boundary
of the covered frequency range. However, in our experience both
approaches have certain drawbacks when analyzing broadband
dielectric spectra, where contributions from EP are observed over
several frequency decades. Instead we found that a distributed RC
equivalent circuit (referred to as DRC circuit henceforth), based on
a Cole-Cole (CC) distribution of relaxation times
\cite{Lunkenheimer1996, Ravaine1973, SANDIFER1974a, MacDonald1976,
Barsoukov2005}, is best suited to describe EP. This is true for such
different materials as aqueous solutions, biological systems,
solid-state electrolytes, ionic liquids and melts, and all kinds of
electronic conductors.

The purpose of the present work is twofold: i.) To provide broadband
experimental spectra with strong EP effects in a variety of very
different materials that may be helpful for future tests of models
of EP (for this purpose, the experimental data are provided in
electronic form in the supplementary information. ii.) To
demonstrate the universal applicability of the DRC circuit for the
modeling of EP effects.




\section{Materials and Methods}
Low-frequency measurements up to $\nu\approx3$~MHz were performed
using frequency-response analyzers (Novocontrol Alpha-A and
Schlumberger 1260 combined with the Chelsea dielectrics interface)
and an autobalance bridge (Hewlett-Packard HP4284A). Here the sample
always was placed in a parallel-plate capacitor.  Data in the higher
frequency ranges were collected using a coaxial reflection
technique, employing the Agilent Impedance/Material Analyzer E4991A
(1~MHz~-~3~GHz) and the autobalance bridge 4294A (40~Hz~-~110~MHz).
Partly, additional measurements beyond 3~GHz were performed via an
open-end coaxial reflection technique and a coaxial transmission
technique using the network analyzers Hewlett-Packard 8510 and
Agilent E8363B. Further details on the experimental techniques can
be found in refs. \cite{Pimenov1996, Wolf2011, Lunkenheimer2000,
Schneider01}. Cooling and heating of the samples were achieved by a
N$_{2}$-gas cryostat (Novocontrol Quatro) and various home-made
ovens.

In the present publication we provide broadband dielectric data for
the following samples (all concentrations in mol percent):

i.) LiBr/water and LiBr/glycerol solutions with the concentrations
$c=0.018$~\% and   $c=0.73$~\%, respectively. For the preparation of
the solutions, 54~wt\% aqueous LiBr solution (Sigma Aldrich),
Glycerol (AnalaR Normapur, 99.5\%), H$_2$O (Merck "Ultrapur"), and
LiBr powder (Sigma Aldrich, purity 99.999\%) were used.

 ii.) LiCl/water solution with $c=17$~\% (Sigma Aldrich, purity $>99$\%).

 iii.) Human whole blood (hematocrit value of 0.39) \cite{Wolf2011}.

 iv.) An aqueous protein solution: lysozyme with
$c=0.009$\% (5~mmol/l) (lysozyme: Sigma Aldrich (Fluka)).

 v.) The glass-forming ionic melt
[Ca(NO${_3})_2$]$_{0.4}$[KNO$_{3}$]$_{0.6}$ (CKN) (see ref.
  \cite{Pimenov1996} for preparation details).

 vi.) The ionic liquid
1-butyl-3-methylimidazolium-bis(trifluoromethylsulfonyl)imide
(BMIM-BMSF) (Iolitec, purity 99\%)

 vii.) Single crystalline La$_{2}$CuO$_{4}$, a parent
compound of high-$T_{c}$ superconductors (see refs.
\cite{LUNKENHEIMER1992, Lunkenheimer1996} for details).

 viii.) Single crystalline La$_{15/8}$Sr$_{1/8}$NiO$_{4}$,
a charge-ordered CDC material \cite{Krohns2009}.

 ix.) Polycrystalline Pr$_{2/3}$Cu$_{3}$Ti$_{4}$O$_{12}$, a
CDC material \cite{Sebald2010}.

\section{Equivalent-circuit models for electrode polarization}

The CPE is often employed to account for EP in ionic conductors
\cite{Bates1988, Macdonald1987, Bordi2001, Fricke1932, Feldman1998,
Bottelberghs1976}. Its admittance is defined by
$Y^{\ast}=A(i\nu)^{n}$ with an exponent $n<1$ and the prefactor $A$.
A theoretical rationale for the CPE can be obtained when considering
a fractional geometry of the electrode/sample interface
\cite{LIU1985, Nyikos1985}. A more intuitive modeling of EP effects
is provided by a parallel connection of a resistance $R_\mathrm{E}$
and a capacitance $C_\mathrm{E}$, connected in series to the bulk
impedance $Z_{\mathrm{B}}$ \cite{Dyre2009, LUNKENHEIMER1992,
Seeger1999, Ritus2002, Schwan1968, Umino2002}. Here $C_\mathrm{E}$
models the high capacitance of the thin blocking layers and
$R_\mathrm{E}$ accounts for their high resistance
($R_\mathrm{E}=\infty$ for complete blocking). This equivalent
circuit leads to a total impedance of

\begin{equation}
\label{equiv}
Z_{\mathrm{total}}=Z_{\mathrm{B}}+\frac{R_{\mathrm{E}}}{1+i\omega
\tau_{\mathrm{E}}}
\end{equation}

\noindent with $\tau_{\mathrm{E}}=R_{\mathrm{E}}C_{\mathrm{E}}$ and
$\omega=2\pi\nu$ the circular frequency. Assuming, e.g., a
frequency-independent bulk resistance and capacitance (i.e., another
RC circuit) for the bulk response, this results in so-called
Maxwell-Wagner relaxation. It leads to a frequency dependence of the
total complex capacitance (and thus of the permittivity
$\varepsilon^*=\varepsilon'-i\varepsilon''$) that is fully
equivalent to the Debye relaxation law \cite{Debye1929} describing
the relaxational response of an ideal dipolar system
\cite{Lunkenheimer2010, Jonscher1983}:

\begin{equation}
\label{debye} \varepsilon^{*}=\varepsilon_{\infty}+
\frac{\Delta\varepsilon}{1+i\omega\tau}-i\frac{\sigma_{dc}}{\omega\varepsilon_{0}}
\end{equation}

\begin{figure}[h!]
\centering
\includegraphics[width=8cm]{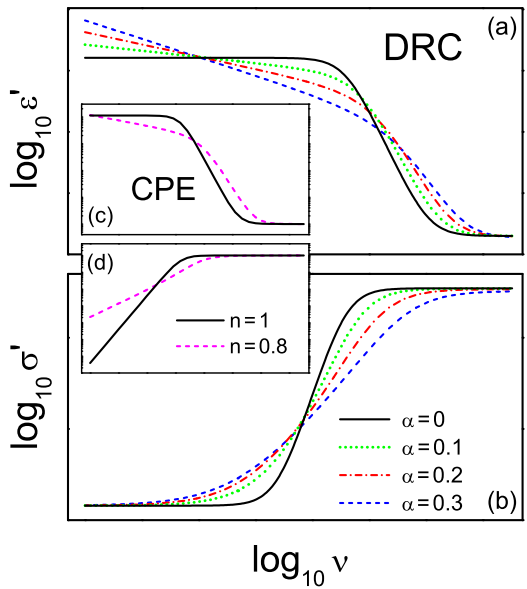}
\caption[x]{\label{fig:simulation} (a) Dielectric constant and (b)
real part of the conductivity calculated according to the
"Cole-Cole"-case ($\beta=1$) of eq. (\ref{eq:RCdist}) for various
values of the distribution parameter $\alpha$. $\alpha=0$
corresponds to the undistributed case, eq. (\ref{equiv}). For the
bulk, frequency independent $\epsilon'$ and $\sigma'$ were assumed,
i.e. a simple parallel RC circuit connected in series to the
impedance of eq. \ref{eq:RCdist}. Frames (c) and (d) show
corresponding curves for a CPE with $n=1$ and $n=0.8$ instead of a
DRC element.}
\end{figure}

\noindent $\tau$ denotes the relaxation time (not identical with
$\tau_{\mathrm{E}}$ \cite{Lunkenheimer2010}) and
$\Delta\varepsilon=\varepsilon_{\mathrm{s}}-\varepsilon_{\infty}$
the dielectric strength. $\varepsilon_{\mathrm{s}}$ and
$\varepsilon_{\infty}$ are the limiting values of the real part of
the permittivity for frequencies well below and above the relaxation
frequency $\nu_{\mathrm{relax}}=1/(2\pi\tau)$, respectively. The
last term in eq. (\ref{debye}) was included to account for the
contribution of dc charge transport to the dielectric loss
$\varepsilon''\propto\sigma'/\nu$, with $\sigma_{dc}$ the dc
conductivity. Typical spectra of $\varepsilon'$ and the conductivity
$\sigma'$ arising from the Maxwell-Wagner relaxation mechanism,
described above, are schematically shown by the solid lines in Figs.
\ref{fig:simulation}(a) and (b), respectively. It should be noted
that in case of the validity of eq. (\ref{equiv}), an intrinsic
dipolar relaxation and a Maxwell-Wagner relaxation caused by EP
cannot be distinguished based on the measured spectra alone.

However, we found that in many cases the above approach does not
lead to a satisfactory description of the experimental data,
especially if the spectra extend to frequencies sufficiently low to
reveal more than just the onset of the EP effects. Then the spectral
features are often found to be more smeared out than expected for
this simple equivalent circuit. Interestingly, intrinsic
relaxations, e.g., in dipolar glass forming liquids
\cite{Lunkenheimer2000}, usually exhibit a similar behavior, i.e. a
broadening of the observed spectral features. In these cases, the
use of an empirical extension of the Debye function, the
Havriliak-Negami function \cite{Havriliak1966}, is well established
to account for the observed deviations from Debye behavior:

\begin{equation}
\label{hn}
\varepsilon^{*}(\nu)=\varepsilon_{\mathrm{\infty}}+\frac{\Delta\varepsilon}{\left[1+(i\omega\tau)^
{1-\alpha}\right]^{\beta}}-i\frac{\sigma_{dc}}{\omega\varepsilon_{0}}
\end{equation}

\noindent The above function with $0\leq\alpha<1$ and $0<\beta\leq1$
includes the special cases with $\beta=1$ and $\alpha=0$, termed
Cole-Cole \cite{Cole1941} and Cole-Davidson \cite{Davidson1950,
Cole1952} function, respectively. These two additional parameters
lead to a broadening of the spectral features, which is commonly
ascribed to a distribution of relaxation times. While the
Havriliak-Negami and the Cole-Davidson functions are purely
empirical, the Cole-Cole distribution of $\tau$ can be approximated
by the microscopic model of Gauss-distributed energy barriers
\cite{Hochli1990}. This formula leads to a symmetric broadening
(compared to the Debye case) of the relaxation peak in
$\varepsilon''$ and, e.g., is often used for the description of
secondary relaxations of glassy matter
\cite{Kudlik1999,Kastner2011}. However, we found that a direct
application of eq. (\ref{hn}) to Maxwell-Wagner relaxations arising
from EP cannot account for the experimental data. An alternative
approach is an extension of eq. (\ref{equiv}) analogous to eq.
(\ref{hn}), arriving at

\begin{equation}
\label{eq:RCdist}
Z_{\mathrm{total}}=Z_{\mathrm{B}}+\frac{R_{\mathrm{E}}}{\left[1+(i\omega\tau_{\mathrm{E}})
^{1-\alpha}\right]^{\beta}}\:
\end{equation}

\noindent with $0\leq\alpha<1$, $0<\beta\leq1$. For the Cole-Cole
case ($\beta=1$), this equation was already considered earlier
\cite{Ravaine1973, SANDIFER1974a, MacDonald1976, Barsoukov2005}. A
generalized Havriliak-Negami version as in eq. (\ref{eq:RCdist}) is
also included in the "LEVM" fit programm by Prof. J. Ross Macdonald
\cite{MacdonaldHome}. However, until now this DRC equivalent circuit
has  only rarely been used for the modeling of EP effects
\cite{Ravaine1973, SANDIFER1974a, Macdonald2007}. As shown in the
further course of the present work, it seems that the CC version of
eq. (\ref{eq:RCdist}) provides good fits of EP-dominated spectra in
a large variety of very different materials.

It has to be stressed that, whereas a simple RC equivalent circuit
(eq. (\ref{equiv})), connected in series to the bulk, produces
permittivity curves identical to a Debye function (eq.
(\ref{debye})), a DRC equivalent circuit based on eq.
(\ref{eq:RCdist}) in general produces permittivity spectra that are
qualitatively different from those of a Havriliak-Negami function,
eq. (\ref{hn}). For the latter, the relaxation time $\tau$ in Eq.
(\ref{hn}) is assumed to be distributed. In the equivalent-circuit
case, the corresponding quantity is $\tau=R_{B}C_{\mathrm{E}}$
($R_{B}$ denotes the bulk resistance), which determines, e.g., the
loss peak position via $\nu_{p}=1/(2\pi\tau)$, just like $\tau$ in
eq. (\ref{hn}) \cite{Lunkenheimer2002, Lunkenheimer2010}. However,
the quantity that is distributed in the equivalent-circuit case is
$\tau_{E}=R_{\mathrm{E}}C_{\mathrm{E}}$, thus leading to different
curve shapes.

Finally it should be mentioned that the DRC circuit for the
Cole-Cole case and a parallel connection of a CPE and a resistor can
produce identical frequency dependences of the dielectric quantities
\cite{Barsoukov2005}. The DRC circuit seems to be the more physical
one because fitting the experimental data with this model provides
direct access to the resistances and capacitances of the blocking
surface layers. However, as will become clear in the further course
of this work, the extended CPE element may find an explanation in a
second, slower hopping transport process within the BE layers
\cite{Serghei2009}.

Figure \ref{fig:simulation} shows typical spectra of $\varepsilon'$
(a) and $\sigma'$ (b) calculated by assuming a DRC equivalent
circuit to account for the EP. For comparison, spectra using a plain
RC circuit (solid lines in (a) and (b)) and a CPE (frames (c) and
(d)), instead of the DRC, are provided. In all cases for the bulk
response a simple parallel RC circuit was assumed, corresponding to
a frequency independent bulk $\varepsilon'$ and $\sigma'$. For the
distribution, the CC case (i.e., $\beta=1$ and $\alpha\neq0$ in eq.
(\ref{eq:RCdist}) was chosen. The curves for the undistributed RC
circuit (solid lines), which are identical to Debye behavior, eq.
(\ref{debye}), show the typical dispersion steps in $\varepsilon'$
and $\sigma'$. With increasing $\alpha$, these steps become smeared
out for both quantities. In addition, the low-frequency plateau of
$\varepsilon'(\nu)$ develops into a weak power law. The low and
high-frequency plateaus of $\sigma'(\nu)$ remain unaffected
(however, we found that for very high values of $\alpha$ the
$\sigma'(\nu)$ curve can cross those limits). Comparing this
behavior with that of the CPE case (Fig. \ref{fig:simulation}(c) and
(d)), the most obvious difference is the missing low-frequency
plateau in $\sigma'(\nu)$ for the latter (d). This would correspond
to completely blocking electrodes. However, adding a parallel
resistance to the CPE leads to such a plateau and, as mentioned
above, in this case both circuits produce identical curves.


In the following, several experimental spectra being governed by
strong EP effects, obtained on partly very different materials, will
be fitted by the models introduced above. All fits were
simultaneously performed for $\varepsilon'(\nu)$ and $\sigma'(\nu)$.
For the experimental data, statistical errors were assumed.

\section{Results and Discussion}

\begin{figure}[h!]
\centering
\includegraphics[width=9.5cm]{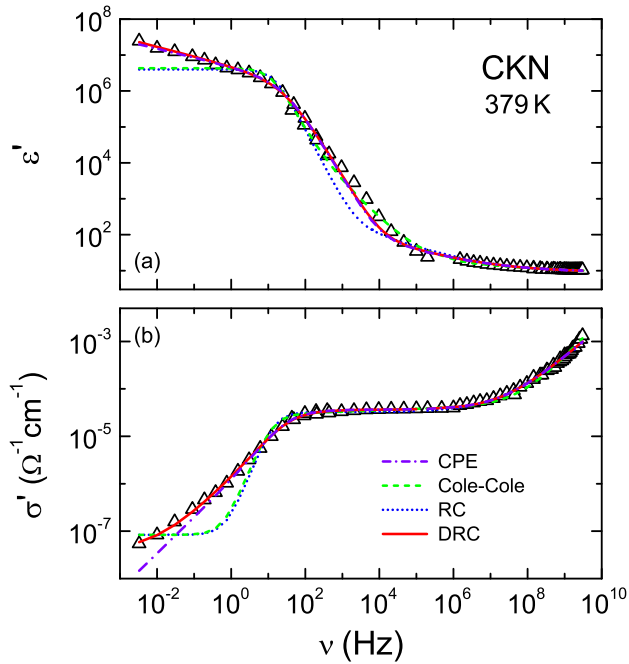}
\caption[x]{\label{fig:vglCKN} (a) Dielectric constant and (b) real
part of the conductivity of CKN at 379~K as function of frequency.
The lines represent fits using different ways to account for the EP
effects: equivalent circuits with a CPE (dash-dotted lines), a
parallel RC circuit, eq. (\ref{equiv}) (dotted lines), or a DRC
circuit, eq. (\ref{eq:RCdist}) with $\beta=1$ (solid lines),
connected in series to the bulk. For the latter, a power-law
increase in $\sigma'(\nu)$ and the corresponding contribution in
$\varepsilon'(\nu)$ was used \cite{Jonscher1983}. For comparison,
the dashed lines show fit curves where the EP effect was treated
like an intrinsic relaxation and fitted by a Cole-Cole function plus
dc-conductivity contribution (eq. (\ref{hn}) with $\beta=1$).}
\end{figure}

Figure \ref{fig:vglCKN} shows broadband spectra of
$\varepsilon'(\nu)$ and $\sigma'(\nu)$ of CKN at 379~K, extending
over 12 frequency decades \cite{Pimenov1996, Lunkenheimer2002b}. In
this supercooled ionic melt, EP effects lead to a strong increase of
the dielectric constant and a strong decrease of the conductivity at
low frequencies. The increase of $\sigma'(\nu)$ at high frequencies
is of intrinsic nature and corresponds to an onset of a
high-frequency minimum in the dielectric loss,
$\varepsilon''\propto\sigma'/\nu$. This typical feature of glass
formers is treated in detail in \cite{Lunkenheimer2002b,
Lunkenheimer1997a}, where data extending to even higher frequencies
are provided. The dotted and dash-dotted lines in Fig.
\ref{fig:vglCKN} show fits with the two standard descriptions of EP,
namely a parallel RC circuit and a CPE, respectively, both connected
in series to the bulk. For the bulk, in addition to the usual
parallel RC circuit, we have assumed a power-law increase in
$\sigma'(\nu)$ and the corresponding contribution in
$\varepsilon'(\nu)$ to phenomenologically account for the mentioned
high-frequency effects \cite{Jonscher1983}. The EP-generated
low-frequency dispersions in $\varepsilon'(\nu)$ and $\sigma'(\nu)$
of CKN are clearly too smeared out to be fitted by a simple RC
circuit (dotted line). In addition, the absence of a clear
low-frequency plateau in $\varepsilon'(\nu)$ is not taken into
account by this approach. Applying this RC equivalent-circuit is
mathematically equivalent to fitting the data with an intrinsic
Debye relaxation, eq. (\ref{debye}). Thus one may expect better fits
if using a broadened function like the Cole-Cole function, eq.
(\ref{hn}) with $\beta=1$. However, as demonstrated by the dashed
lines in Fig. \ref{fig:vglCKN}, such fits are still unsatisfactory.
This is also the case for the Havriliak-Negami and Cole-Davidson
variant of eq. (\ref{hn}) (not shown).

While the CPE (dash-dotted lines in Fig. \ref{fig:vglCKN}) does a
good job in the description of $\varepsilon'(\nu)$, it fails to
account for the onset of a low-frequency plateau observed in
$\sigma'(\nu)$ at $\nu<0.1$~Hz. This plateau, which is often
overlooked because of a limited frequency range or low ion mobility,
obviously is due to an incomplete blocking of the electrodes. It
indicates a second conductivity mechanism that starts to prevail at
$\nu<10^{-1}$~Hz when the main charge-transport mechanism is
blocked. It could be due to electronic conductivity or to a second
ion species with much lower mobility, which is not yet blocked at
the lowest frequencies investigated \cite{Sanabria2006}. However, it
seems unlikely that the mobilities of the different ion species
(K$^{+}$, Ca$^{2+}$, and NO$_{3}^{-}$ in the present case) differ by
about three orders of magnitude as suggested by a comparison of the
plateaus at $\nu\rightarrow0$ and $\nu\approx10^{4}$~Hz
\cite{Ribeiro2004}. Another possibility is that the ions located
within the space charge region, close to the electrode, still have
some mobility left. In line with the model proposed in ref.
\cite{Serghei2009}, this mobility can be assumed to be much lower
than in the bulk, but not zero, thus explaining the approach of an
additional low-frequency plateau in $\sigma'(\nu)$. Then it seems
plausible that at extremely low frequencies a second, final blocking
effect should arise, i.e. $\sigma'(\nu)$ should decrease again when
the ions within the space charge region finally reach the sample
boundary. As shown in the further course of this work, this is
indeed observed in several of the investigated materials.

As demonstrated by the solid lines in Fig. \ref{fig:vglCKN},
modeling the BEs by a DRC circuit yields satisfying fitting curves
of $\varepsilon'(\nu)$ and $\sigma'(\nu)$ in the whole frequency
range, including the onset of a low-frequency plateau in the
conductivity. The resistance arising from the second
charge-transport mechanism is taken into account by the resistor
within the parallel RC equivalent circuit, modeling the EP effects.
We find $\alpha=0.26$ and $\beta=1$, indicating a moderate,
symmetrical distribution of relaxation times $\tau_{E}$ (Cole-Cole
case). Broadband spectra obtained at other temperatures in CKN
\cite{Pimenov1996, Lunkenheimer2002b} also can be well modeled in
this way. In fact, in an earlier work \cite{Macdonald2007} a
sophisticated model that included a DRC circuit (termed "ZC" there)
to account for EP, was used to obtain excellent fits of spectra in
CKN at four temperatures between 342 and 361K.

\begin{figure}[h!]
\centering
\includegraphics[width=9.5cm]{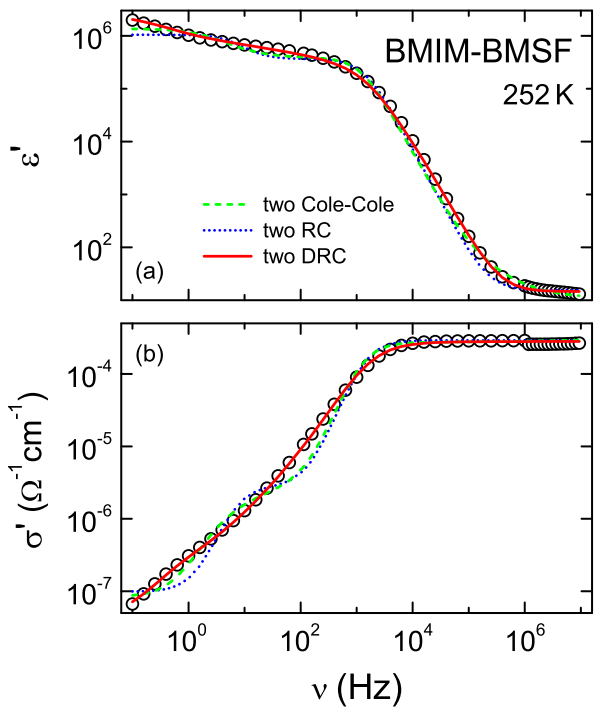}
\caption[x]{\label{fig:vglBB} (a) Dielectric constant and (b) real
part of the conductivity of BMIM-BMSF at 252~K as function of
frequency. The lines represent fits assuming different models for
the EP effects: the sum of two Cole-Cole functions, eq. (\ref{hn})
with $\beta=1$ (dashed lines), two parallel RC circuits, eq.
(\ref{equiv}), connected in series to the bulk (dotted lines), and
two DRC circuits, eq. (\ref{eq:RCdist}) with $\beta=1$, also
connected in series to the bulk impedance (solid lines). For the
latter two cases, the bulk is modeled by a parallel RC circuit.}
\end{figure}

Figure \ref{fig:vglBB} provides results on BMIM-BMSF, an ionic
liquid, i.e. a substance solely consisting of cations and anions
that is liquid at room temperature. This class of materials has
attracted considerable interest in recent years due to their
possible technical relevance as environment-friendly solvents and as
new types of electrolytes used, e.g., in batteries \cite{Welton1999,
Armand2009}. Again, typical EP effects are observed in
$\varepsilon'(\nu)$ (a) and $\sigma'(\nu)$ (b). The upward bending
of the $\varepsilon'(\nu)$ curve in the upper-plateau region of the
main dispersion step below $\nu\approx10$~Hz clearly indicates the
presence of a second dispersion step. In Fig. \ref{fig:vglBB}(b),
after the initial decrease of $\sigma'(\nu)$  below about
$3\times10^{3}$~Hz, one may suspect the approach of a second plateau
at around $10$~Hz with amplitude of
$\sim10^{-6}~\Omega^{-1}\textrm{cm}^{-1}$, which is followed by a
second decrease below about 1~Hz. As discussed above, the latter
finding may well indicate the final blocking of a second
charge-transport mechanism. Admittedly, the mentioned features are
not well pronounced and rather smeared out. However, we found that
fits with a single DRC element are not able to provide a good
description of the present data. Instead, the experimental spectra
can be very well fitted by assuming \emph{two} DRC elements in
series to the bulk (solid lines) modeling the blocking of the two
transport mechanisms. Here the bulk is assumed to have
frequency-independent $\varepsilon'$ and $\sigma'$, i.e. it is
modeled by another (undistributed) parallel RC circuit. From the
fits, we obtain $\alpha=0.18$ and $\beta=1$ for the DRC element,
modeling the main dispersion step. For the low-frequency one, a pure
RC element ($\alpha=0$, $\beta=1$) was sufficient, which may be due
to the fact that it is only partly visible in the investigated
frequency range. As indicated by the dotted lines in Fig.
\ref{fig:vglBB}, a combination of two undistributed RC circuits (eq.
(\ref{equiv})) in series to the bulk, cannot explain the observed
frequency dependence. Using instead the sum of two Cole-Cole
functions (eq. (\ref{hn}) with $\beta=1$; dashed lines), i.e.
treating the data as arising from two intrinsic dipolar relaxations
with different time constants, also does not account for the
experimental data.

\begin{figure}[h!]
\centering
\includegraphics[width=9.5cm]{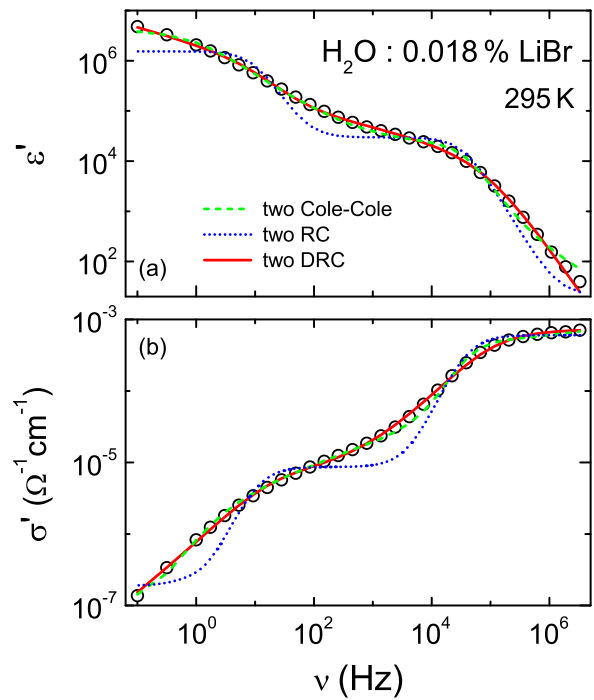}
\caption[x]{\label{fig:vglLiBr} (a) Dielectric constant and (b)
conductivity of a 0.018~\% LiBr/water solution at 295~K as function
of frequency. The meanings of the lines are the same as in Fig.
\ref{fig:vglBB}.}
\end{figure}

As an example for a classical electrolyte solution, Fig.
\ref{fig:vglLiBr} provides the frequency dependence of
$\varepsilon'$ and $\sigma'$ of a 0.018~\% LiBr/water solution at
295~K. The real part of the permittivity (a) reveals two smeared-out
dispersion steps with points of inflection at around 10 and
$10^5$~Hz, which are caused by EP effects. The conductivity (b) also
shows two dispersion features. The high-frequency plateau in
$\sigma'(\nu)$ of about $10^{-3}~\Omega^{-1}\textrm{cm}^{-1}$
corresponds to the intrinsic ion conductivity of this solution,
dominated by the main charge-transport process (see discussion of
the CKN data). This system exhibits a high conductivity and an onset
of the EP effects at rather high frequencies (about $10^{5}$~Hz in
$\sigma'(\nu)$, compared to $3\times10^{3}$~Hz in BMIM-BMSF at 252~K
and $10^{2}$~Hz in CKN at 379~K, cf. Figs. \ref{fig:vglBB} and
\ref{fig:vglCKN}). Probably for this reason, the appearance of a
second plateau (at about $10^{2}$~Hz and
$\sim10^{-5}~\Omega^{-1}\textrm{cm}^{-1}$) is well pronounced in
this electrolyte and, in addition, the final blocking is well seen
(further decrease of $\sigma'(\nu)$ below $\sim10$~Hz). Both effects
clearly indicate a second conductivity mechanism that becomes
blocked at lower frequencies, as already suspected in the discussion
of the CKN and BMIM-BMSF data.

The lines in Fig. \ref{fig:vglLiBr} again are fits with different
empirical functions to account for the BEs. As demonstrated by the
dotted lines, two parallel RC equivalent circuits (eq.
(\ref{equiv})), both connected in series to the bulk impedance,
obviously cannot account for the experimental data. Here again the
bulk was assumed to have frequency-independent $\varepsilon'$ and
$\sigma'$. The sum of two Cole-Cole functions (eq. (\ref{hn}) with
$\beta=1$) leads to much better, but still not perfect fits (dashed
lines). In contrast, assuming two DRC elements, eq.
(\ref{eq:RCdist}), with $\beta=1$, connected in series to the bulk
(solid line), provides nearly perfect fits. The values of the
distribution parameter $\alpha$ are 0.32 (for the low frequency EP
effect) and 0.28 (for the high frequency EP effect).

\begin{figure}[h!]
\centering
\includegraphics[width=8.5cm]{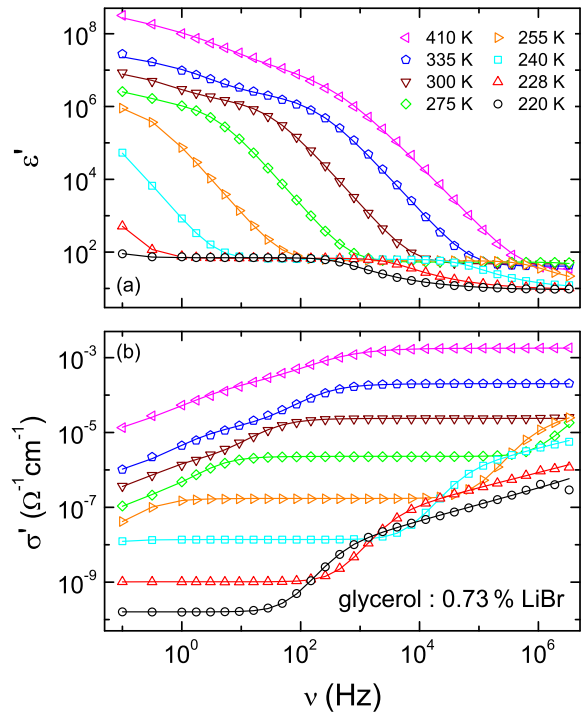}
\caption[x]{\label{fig:GlyLiBr} (a) Dielectric constant and (b)
conductivity of a 0.73~\% LiBr/glycerol solution at different
temperatures as function of frequency. The lines are fits with two
DRC circuits, eq. (\ref{eq:RCdist}) with $\beta=1$, in series to the
bulk to account for the EP. For the bulk, an intrinsic dipolar
relaxation, described by the Cole-Davidson formula and a
frequency-independent dc conductivity were used (eq. (\ref{hn}) with
$\alpha=0$).}
\end{figure}

As an example for the temperature-dependent development of EP
effects, Fig. \ref{fig:GlyLiBr} shows  $\varepsilon'(\nu)$ (a) and
$\sigma'(\nu)$ (b) of a 0.73\% LiBr/glycerol solution for various
temperatures between 220~K and 410~K. For the lowest temperature
shown (220~K) only the intrinsic $\alpha$-relaxation of glycerol
(cooperative reorientation of the dipolar glycerol molecules
\cite{Lunkenheimer2000}) and the bulk conductivity arising from the
ionic charge transport can be seen in this frequency range. This
relaxation is characterized by steps in $\varepsilon'(\nu)$ and in
$\sigma'(\nu)$ (the latter corresponding to the well-known loss
peaks in $\varepsilon''(\nu)$ \cite{Lunkenheimer2000}). The bulk
conductivity leads to the low-frequency plateau in $\sigma'(\nu)$ of
about $10^{-10}~\Omega^{-1}\textrm{cm}^{-1}$. Due to the low ion
mobility at this temperature, the EP effects lie outside the
investigated frequency range. With increasing temperature, the
mobility rises and the EP shows up as a strong increase in
$\varepsilon'(\nu)$ at low frequencies that finally develops into a
steplike curve. In addition, a decrease in the conductivity for
decreasing frequency arises above 240~K. At even higher
temperatures, a second, smeared-out step is revealed in
$\varepsilon'(\nu)$, accompanied by a weak corresponding feature in
$\sigma'(\nu)$. Similar to the behavior in the aqueous LiBr solution
(Fig. \ref{fig:vglLiBr}), this further dispersion indicates a second
charge-transport mechanism as discussed above. As demonstrated by
the lines in Fig. \ref{fig:GlyLiBr}, perfect fits of all these
spectra are achieved by assuming a DRC circuit (eq.
(\ref{eq:RCdist})) with $\beta=1$ for the description of the
observed EP effects, connected in series to the bulk response. For
the latter, the $\alpha$-relaxation is accounted for by the
Cole-Davidson function, eq. (\ref{hn}) with $\alpha=0$
\cite{Lunkenheimer2000, Kohler2008}. The distribution parameters
$\alpha$ of the slower EP effect vary between 0.1 and 0.33; for the
second one we obtained values between 0.1 and 0.29.

\begin{figure}[h!]
\centering
\includegraphics[width=9.5cm]{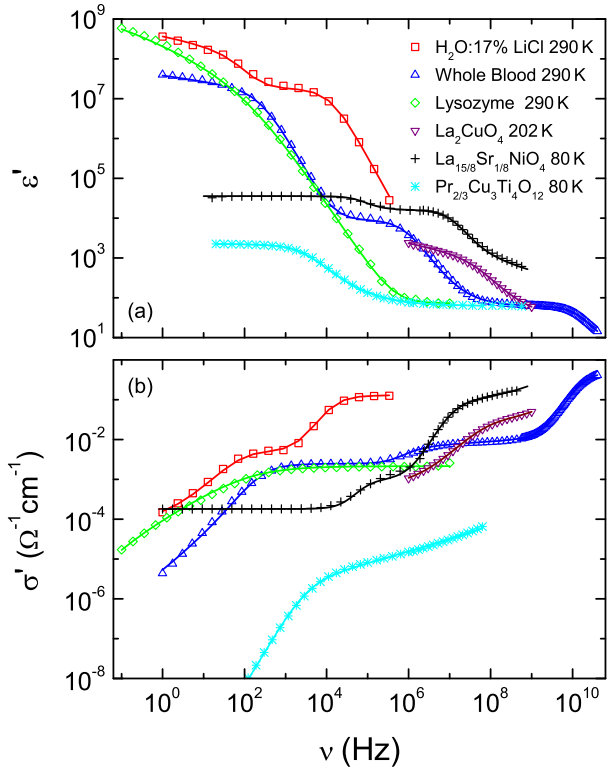}
\caption[x]{\label{fig:final-many} (a) Dielectric constant and (b)
real part of the conductivity of different samples that are affected
by EP effects: 17\% LiCl/water solution at 290~K (squares), human
whole blood at 290~K (triangles), 0.009\% Lysozym/water solution
(diamonds), single-crystalline La$_{2}$CuO$_{4}$ (down triangles),
single crystalline La$_{15/8}$Sr$_{1/8}$NiO$_{4}$ (crosses), and
polycrystalline Pr$_{2/3}$Cu$_{3}$Ti$_{4}$O$_{12}$ (stars). The
lines are fits with one or two DRC circuits, connected in series to
the bulk, which is modeled by a parallel RC circuit.}
\end{figure}

Further examples of dielectric spectra, dominated by EP are provided
in Fig. \ref{fig:final-many}. It shows the dielectric properties of
a variety of partly very different materials, including a highly
concentrated salt solution, two biological samples, and three
electronic conductors. The squares represent the results obtained on
a 17\% LiCl/water solution at 290~K. Its frequency spectra reveal
two strong dispersion steps in $\varepsilon'(\nu)$ (a) and
$\sigma'(\nu)$ (b). As demonstrated by the lines, the spectra are
well fitted with two Cole-Cole DRC elements in series to the sample
impedance, which is modeled by a simple RC circuit. The broadening
of the low-frequency dispersion step is characterized by
$\alpha=0.22$; the faster one could be modeled with a pure RC
equivalent circuit (i.e. $\alpha=0$, $\beta=1$).

The triangles in Fig. \ref{fig:final-many} represent the broadband
data of human blood at 290~K \cite{Wolf2011}. Blood can be
considered as an aqueous solution of a variety of different salts
with the addition of a considerable fraction of suspended
"particles", mainly red blood cells. As treated in detail in ref.
\cite{Wolf2011}, at $\nu>10^{5}$~Hz, those data are dominated by the
well-known $\beta$- and $\gamma$-dispersions of blood
\cite{Wolf2011, Schwan1957}. The $\beta$-dispersion (at about
$10^{6}$ - $10^{7}$~Hz) arises from a Maxwell-Wagner relaxation
caused by the membrane of the red blood cells \cite{Schwan1957}. The
$\gamma$-dispersion (at about $10^{10}$~Hz) is caused by the
cooperative reorientational motion of water molecules
\cite{Wolf2011, Schwan1957}. At $\nu<10^{5}$~Hz, strong EP effects
show up in the blood spectra of Fig. \ref{fig:final-many}. The
complete broadband spectra are well fitted with a single Cole-Cole
DRC circuit (eq. (\ref{eq:RCdist}) with $\beta=1$), connected in
series to the bulk impedance. The latter is modeled by the sum of
two "normal" Cole-Cole functions (i.e. they are defined in
$\varepsilon^*$; eq. (\ref{hn}) with $\beta=1$) to account for the
$\beta$- and $\gamma$-dispersions. The finding that a single DRC
circuit is sufficient to account for the data may be ascribed to the
fact that the onset of the EP effects in blood occurs at rather low
frequencies, comparable to the results in CKN (Fig.
\ref{fig:vglCKN}).

The diamonds in Fig. \ref{fig:final-many} show the results for a
0.009\% (5~mmol/l) Lysozyme solution. Protein solutions usually
contain free ions and indeed EP effects show up in the present
spectra, leading to strong dispersion in both quantities. The very
gradual transition to a low-frequency plateau in $\varepsilon'(\nu)$
at $\nu<10^{2}$~Hz indicates the presence of a second, smaller step
on top of the main dispersion. Indeed, excellent fits of the
experimental spectra were achieved by using two DRC elements in
series to the bulk. The DRC circuit leading to the smaller,
low-frequency step is Debye-like ($\alpha=0$, $\beta=1$); the broad
main dispersion step is characterized by $\alpha=0.46$ and
$\beta=1$.

Finally, Fig. \ref{fig:final-many} also provides some typical data
on electronically conducting materials, where the EP arises from the
formation of Schottky diodes at the electrode/sample interface. Even
when the polarity of the oscillating field changes, one of the two
oppositely poled diodes at both sides of the sample always is
blocked and dominates the response at low frequencies. In recent
years, especially EP effects in non-metallic transition-metal oxides
have gained increasing significance: Many of these materials show
effects like magnetocapacitance, multiferroicity or the occurrence
of very large dielectric constants that have triggered numerous
dielectric investigations \cite{Lunkenheimer2010, Spaldin2005,
Cheong2007}. However, EP effects can considerably hamper the
detection of their intrinsic dielectric properties and even be
misinterpreted as being of intrinsic origin \cite{Lunkenheimer2002,
Lunkenheimer2004, Catalan2006, Loidl2008}. The examples provided in
Fig. \ref{fig:final-many} include single-crystalline
La$_{2}$CuO$_{4}$, a parent compound of high-$T_{c}$ superconductors
\cite{LUNKENHEIMER1992, Lunkenheimer1996}, single crystalline
La$_{15/8}$Sr$_{1/8}$NiO$_{4}$, a charge-ordered CDC material
\cite{Krohns2009}, and polycrystalline
Pr$_{2/3}$Cu$_{3}$Ti$_{4}$O$_{12}$ \cite{Sebald2010}, a material
structurally closely related to the well-known CDC material
CaCu$_{3}$Ti$_{4}$O$_{12}$ \cite{Homes2001}. For all these
materials, strong EP effects are revealed in Fig.
\ref{fig:final-many}, which look qualitatively similar to those in
the ionic conductors. In all cases, the spectra can well be fitted
by using a DRC circuit to account for the EP (lines). For
La$_{15/8}$Sr$_{1/8}$NiO$_{4}$ a second DRC circuit was used to
cover the second dispersion step seen below about $10^{6}$~Hz. In
ref. \cite{Krohns2009}, by measurements with different contact types
the high-frequency step was clearly identified to arise from EP. The
origin of the low-frequency step is not clarified yet but it was
speculated that it may be connected to the inhomogeneous charge
distribution in this prototypical charge-ordered material
\cite{Krohns2009}. We also found the DRC to be applicable for the
description of EP effects in Cu$_{2}$Ta$_{4}$O$_{12}$
\cite{Lunkenheimer2004} and CaCu$_{3}$Ti$_{4}$O$_{12}$, another CDC
material \cite{Renner2004}.

\section{Summary and Conclusions}

In the present work, we have provided dielectric spectra showing
strong EP effects measured in a variety of materials. They partly
belong to very different material classes like salt solutions, ionic
liquids and melts, biological materials, and electronic
semiconductors. Our results are intended to provide a comprehensive
data base for comparing current and future approaches for the
modeling of EP effects in dielectric spectra. For this purpose, the
data are made available in electronic form in the supplementary
information. In addition, in the present work we have demonstrated
that the DRC equivalent circuit is an easy-to-apply phenomenological
model for fitting EP effects in dielectric spectra, which works well
for all the different material classes investigated. However, this
model certainly cannot be regarded as an alternative to microscopic
models as discussed, e.g., in refs. \cite{Bazant2004, Sanabria2006,
Serghei2009, Macdonald2010, Macdonald2011,  Bazant2011}. On the
other hand, it has proven useful for a quick, phenomenological
description of EP effects, enabling fits of experimental data in the
complete investigated frequency range, which can help in the
determination of the intrinsic bulk properties of the investigated
material.

In all the investigated ionically conducting materials, we find
evidence for a second charge-transport process that leads to a
second plateau in the frequency-dependent conductivity at
frequencies below the commonly observed decrease of $\sigma'(\nu)$
due to blocking. In addition, in most of these materials we find a
further decrease of $\sigma'(\nu)$ at the lowest frequencies
investigated, indicating final blocking of the charge carriers
involved in this second process. These seemingly quite universal
features have been only rarely detected so far: In most works on
ionic conductors only the onset of the EP effects is observed in the
spectra, and the data can be fitted by simple models like an
undistributed RC circuit or a CPE. It is unlikely that the second
dispersion, observed in measurements extending to sufficiently low
frequencies, is due to a second, much slower ion species or due to
electronic conduction. Instead, to us an explanation in line with
the model proposed in ref. \cite{Serghei2009} seems more likely: The
same ion species moves fast in the bulk region and much slower in
the space-charge region, thus leading to the observed two dispersion
effects. In ref. \cite{Serghei2009}, a specific hopping-conductivity
model \cite{DYRE1985} was used to model the transport process in the
region close to the electrode. Here we use the DRC circuit with
$\beta=1$, which is equivalent to a CPE connected in parallel to a
resistor \cite{Barsoukov2005}. A CPE, sometimes termed "universal
dielectric response" \cite{Jonscher1983}, is often employed to
describe hopping conductivity in various materials and indeed it
leads to very similar frequency dependence as the model considered
in \cite{Serghei2009}. Thus, using the DRC circuit is in line with
the model of ref. \cite{Serghei2009} and adding a second DRC element
also accounts for the final blocking of the slow ions in the BE
region.

It is an interesting finding that the DRC element also provides a
perfect description of the EP effects observed in electronic
conductors.  In fact, La$_{2}$CuO$_{4}$ (Fig. \ref{fig:final-many})
was the first material, where we introduced this circuit for the
description of an electronic conductor 15 years ago
\cite{Lunkenheimer1996}. Great care has to be taken to avoid the
misinterpretation of relaxation-like features in the spectra of
semiconducting materials as intrinsic effects that could be
indicative, e.g., of ferroelectricity or "colossal" dielectric
constants. Thus the correct modeling of these effects is essential.
The fact that the present DRC-circuit approach leads to a different
frequency dependence than an intrinsic relaxation process, modeled,
e.g., by a Cole-Cole function (see, e.g., Fig. \ref{fig:vglBB}),
provides some possibility to distinguish between intrinsic and
extrinsic effects. In addition, of course the procedures proposed in
ref. \cite{Lunkenheimer2002} (e.g., performing measurements with
different contact types, etc.) should be performed to arrive at a
definite conclusion.

\bigskip
\bigskip

This work was partly supported by the Deutsche
Forschungsgemeinschaft via Research Unit FOR 1394 and via the
Transregional Collaborative Research Center TRR 80. We thank A.
F\"{a}lschle for performing the measurements in BMIM-BMSF.



\bibliographystyle{epj}
\bibliography{RCpaper}







\end{document}